\documentclass{konferencja}

\usepackage{graphicx}

\title{How to simulate the Universe in a box? Testing modified gravity with a N-body simulation code.\\
\small To appear in Proceedings of the Science-Technology Conference for phd students and young scientists - 
\textit{"Young scientist considering the challenges of modern technology", 22-24 September, 2008, Warsaw}
}

\author{Wojciech A. Hellwing \email{pchela@camk.edu.pl} \affiliation{Nicolaus Copernicus Astronomical Center, Bartycka 18, 00-716 Warsaw, Poland}}

\begin{document}

\maketitle

\begin{abstract}

In this paper we present preliminary results from cosmological simulations 
of modified gravity in the dark matter sector. 
Our results show improvements over
standard cold dark matter cosmology. The abundance of low-mass haloes in the modified gravity model 
fit observations better than the conventional theory, while the differences of the modified density fluctuation power spectrum differs from the standard,
$\Lambda$CDM power spectrum are small enough to make these two models observationally indistinguishable at large scales.   

\end{abstract}

\section{Introduction}
Over the past 30 years cosmology became a successful and mature empirical 
science. The growing accuracy of astronomical observations gave the possibility to test theoretical 
models. Past decade has established the "standard cosmological model" dubbed also as the 
$\Lambda$CDM cosmology, or the "concordance model". In this picture the evolution of the 
expanding universe, dominated by non-relativistic (cold) dark mater particles 
is governed by Einstein's gravity with a positive cosmological constant. Although the 
standard model is very successful in explaining many observations it still has some 
difficulties. The physical nature of the dark matter (DM) sector is still unknown. 
Moreover, the mass distribution in the real universe (in particular, the emptiness of voids)
differs significantly from the predictions of the $\Lambda$CDM model.
Modifications of the standard gravity theory have 
been proposed to deal with this impasse. In this paper we show preliminary 
results of our work based on one of such modifications. 

\section{Long-range scalar interactions in CDM}

The conventional cosmological paradigm is the \textbf{$\Lambda$CDM} model. The model 
consists of a set of assumptions and 'best fit' parameters chosen to describe 
a certain set of astronomical observations.
Evolution of the Universe as whole, when we think of scales of order of a few Gpc\footnote{1 Gigaparsec = $10^9$ 
parsecs (pc), 1 pc = 3.26 light years} is governed by Einstein's general relativity theory. This evolution 
obeys Friedman equations. In $\Lambda$CDM the Universe is 
undergoing a phase of accelerated expansion due to large content of Dark Energy, a mysterious phenomenon of 
unknown physical nature that drive global repulsion. On smaller scales (of order of hundredths Mpc) we can describe
the Universe using Newtonian mechanics.
This model is very successful on large scales. It agrees well with observations of Cosmic Microwave 
Background Radiation (CMBR), dynamics of clusters of galaxies, and provides a natural
explanation for the origin and the observed abundances of light elements. Although 
the power spectrum of the CMBR angular fluctuations, predicted by the $\Lambda$CDM model agrees with 
observations (cf. \textit{Wilkinson Microwave Anisotropy Probe} WMAP results \cite{WMAP}), while the
power spectrum of density fluctuations agrees with the large-scale clustering seen in galaxy surveys,  
the {$\Lambda$}CDM N-body simulations are in contradiction with observations on small scales.

\subsection{Shortcomings of $\Lambda$CDM cosmology}
A thorough discussion of all of the difficulties faced by the  $\Lambda$CDM is beyond the scope of this paper.
Instead, we shall briefly 
point out the most important  issues, and in the next section we propose 
a possible solution for them. The problems we address are:
\begin{enumerate}
 
 \item High overabundance of satellites and small galactic haloes in N-body simulations, which 
 disagree with Local Group observations,
 
 \item $\Lambda$CDM simulations suggest that intergalactic and inter-cluster voids are 
 not so quite empty. There are still lots of  'debris' - unaccreted matter left in 
 simulated voids. Observations favor an opposite picture, where voids are really empty of bright galaxies and matter,
 
 \item  The problem of steep central density distributions of galaxy haloes. Simulations favor a cusp with a logarithmic slope for density distribution, whereas observations suggest haloes with a core of constant density (see e.g. \cite{Power2003}, \cite{Zackrisson}),

 \item N-body simulations present a picture of high accretion activity in the recent past. 
 This poses a real problem to galaxy formation and evolution theories. Large spiral galaxies 
 would be ``dimmed'' and their spiral structure probably destroyed by late time merging and accretion activity,
 contrary to observations.
 \end{enumerate}

\subsection{Scalar ,,enhanced'' gravity}

Long-range scalar interactions in the DM sector was proposed by P.~J.~E. Peebles 
\& S.~Gubser as a possible solution to pure $\Lambda$CDM cosmology problems (c.f. \cite{Peebles1}, 
\cite{Peebles2}). Physical motivation for this model comes from the string theory. We consider DM particles as strings. 
Such objects will interacts by gravitation and additional long-range force comes from exchange of a massless scalar. 
A screening length for scalar interaction is dynamically generated by the presence of light particles with a Yukawa-like 
coupling to the scalar field. In this paper we consider one species of strings as DM particles 
with a force law rising from the potential of a form:

\begin{equation}
\label{scalar-pot-r}
\phi(\vec{r})=-G\int d^3\vec{r'}{\rho(\vec{r'})\over|\vec{r}-\vec{r'}|}\left(1+\beta e^{-|\vec{r}-\vec{r'}|\over r_s}\right)\,\,,
\end{equation}
where $\beta$ is a "scalar to gravity" ratio parameter and $r_s$ is the screening length. $\beta$ is a measure of the power of 
scalar interaction compared to gravity interaction between two DM particles. The screening length parameter due to mechanism of 
dynamical screening is constant in co-moving coordinates. In Fourier space this eq. transforms to

\begin{equation}
\label{scalar-pot}
\widetilde{\nabla^2\phi(\vec{k})}={3\over2}{H_0^2\Omega_0\over a}\widetilde{\delta(\vec{k})}\left[1+{\beta\over 1+(\vec{k}\cdot r_s)^{-2}}\right]\,\,,
\end{equation}
where $\widetilde{\phi(\vec{k})}$ is the Fourier transformed potential, $H_0$ is present Hubble constant value, $\Omega_{0}$ 
is dimensionless parameter of present matter density contribution and $\widetilde{\delta(\vec{k})}$ is Fourier transformed 
density contrast. As we can see mechanism of screening leads to screening of scalar force on a particular Fourier modes during 
time evolution. This gives us limiting behavior on large and small scales:
\begin{eqnarray}
\phi\rightarrow (1+\beta)\phi_{Newton}\quad {\rm for}\;\; r\ll r_s,\\
\phi\rightarrow \phi_{Newton}\quad {\rm for}\;\; r\gg r_s,
\end{eqnarray}
Considerations on the grounds of the string theory suggest that sensible values of two free parameters of 
this model are $\beta\sim 1$ and $r_s\sim 1$Mpc. Earlier numerical and analytical study of this model (see \cite{Peebles3}) 
suggest that presence of the scalar interaction will result in earlier formation of structure on small-scales and more rapid 
pace of formation of structures. Other expected effects are the absence of low mass halos in  voids and a lower abundance of 
sub-galactic halos. In the light of presented difficulties of the $\Lambda$CDM scenario such properties would be most
welcome.

\section{N-body simulations - evolving universe in your computer}

One of the tools of choice for modern theoretical cosmology is a collisionless N-body simulation.
Computer N-body codes provide a way to investigate non-linear gravitational evolution of complex particle systems. 
There are many variety of N-body codes,
algorithms and techniques, but all of them essentially model evolution by following the trajectories of particles under 
their mutual gravity.

In cosmology one is using N-body codes to model the growth of structures in time from initial, small-amplitude 
inhomogeneities into the highly clustered structure, observed today. The quality of an N-body code  
rests on (1) generating accurate initial conditions (IC) i.e. initial positions 
and velocities of particles, and (2) as precise, as possible solution of the equations of motion 
to find particle trajectories.

Using measured power spectrum of CMBR photons we can calculate IC for an N-body experiment at sufficiently 
early time of cosmic evolution. Commonly used scheme of generating IC applies the Zeldovich approximation to move 
particles from a Lagrangian point $\vec{q}$ to a Eulerian point $\vec{x}$(e.g. \cite{Knebe_1_2005}, \cite{Efstathiou85}).

Having set up the IC we can start to model dynamics of a collisionless system representing the Dark Matter particles 
in the Universe. To do that a N-body code is solving simultaneously the collisionless Boltzmann equation (CBE):
\begin{equation}
{\partial f\over\partial t}+\sum_{i=1}^3\left(v_i{\partial f\over\partial x_i}-{\partial\Phi\over\partial x_i}{\partial f\over\partial v_i}=0\right)
\label{CBE}
\end{equation}
and the Poisson equation:
\begin{equation}
 \nabla^2\Phi(\vec{r})=4\pi G\rho(\vec{r})
\label{Poisson} 
\end{equation}
Eq.(\ref{CBE}) is solved using method of characteristics (see \cite{LCB} for details). 
The trajectories we get by time integrating $N$ points $\{\vec{r_i}(t),\vec{v}(t)\}$ sampled form the 
distribution function $f$ at time $t_{ini}$ form a representative sample of $f$ at any time $t$ since $f$ 
is constant along any given trajectory. The main job to be done in N-body code is then to solve Eq. (\ref{Poisson}) 
for a given particle system and move the particles to the next time frame according to the equations of motions.

In this study we are using a \textbf{Particle Mesh Code} (PM) (see \cite{Efstathiou85}, \cite{LCB}) that allows 
one to use a N-body simulation code to follow evolution of a system containing millions of particles. 

\subsection{The simulation code - AMIGA}
We use a modified AMIGA code (\textit{Adaptive Mesh Investigations of Galaxy Assembly}), 
a \verb#c# code for cosmological simulations. AMIGA is a successor of the MLAPM code (\textit{Multi-Level Adaptive 
Particle Mesh}) (Knebe, Green and Binney \cite{MLAPM_paper}) and it is purely grid-based N-body code. Due to large number of simulation runs we had to perform, we used a pure PM part 
of AMIGA code to greatly shorten the time of integration at the expense of force resolution.

To enable AMIGA to do modified gravity simulations, we changed the Green's function form to account for investigated effects. 
The 3-D discrete Poisson equation for standard gravity in Fourier space reads
\begin{equation}
 \widetilde{\phi(k)} = {3\over 2}{H_0^2\Omega_{0}\over a}\widetilde{\delta(k)}\times G_k,
\end{equation}
where $G_k$ is Green's function:
\begin{equation}
G(k_x,k_y,k_z) = -\pi / \left\{ NG^2 \left[ \sin^2 (\pi k_x / NG) + \sin^2 (\pi k_y / NG ) + \sin^2 (\pi k_z / NG) \right]\right\} \, .
\end{equation}
Using Eq. (\ref{scalar-pot}) we modify Green function to get proper potential in scalar-interaction model:
\begin{equation}
G_k^{\rm scalar} = G_k^{\rm Newton}\times\left(1+{\beta\over 1 + (k\cdot r_s)^{-2}}\right)
\end{equation}

We use above modification together with the PM part of the AMIGA code to evolve particle system from early 
epoch to present time of cosmic evolution. Simulations results allow us to study the impact of scalar 
interactions on formation of large-scale structures in the Universe.

\section{Simulations results}
To study the impact made by long-range scalar interaction on process of structure formation we have 
performed a series of numerical cosmological simulations. IC for our simulation runs were generated 
using Klypin \& Holtzman \verb#PMcode# \cite{KlypinPM}. We use adiabatic IC with truncated CDM power 
spectrum generated with a help of \verb#cmbfast# code for line-of-sight integration of CMBR (see \cite{cmbfast}). 
In Table \ref{table1} we present values of cosmological parameters used in our simulations. We have tested 
parameter space for $\beta = -0.5, 0.0, 0.2, 0.5, 0.7, 0.8$ and $r_s = 0.5, 1.0, 2.0, 5.0$ Mpc for simulations 
of $123^3$ particles over $256^3$ mesh in a box size $200 h^{-1}$Mpc.
\begin{table}
\caption{Cosmological parameters used in simulation runs}\label{table1}
\centering
\begin{tabular*}{\textwidth}{@{\extracolsep{\fill}}|c|c|c|c|c|c|}
\hline
Parameter & $\sigma_8$ & n & $\Omega_0$ & $\Omega_{\lambda}$ & h \\ \hline
value & 0.8 & 1 & 0.3 & 0.7 & 0.7\\ \hline
description & \parbox{2.4cm}{P(k)-normalization} & \parbox{1.8cm}{spectrum\\ scalar index} &\parbox{1.8cm}{matter density\\ at $a=1$} & \parbox{2.5cm}{dark energy\\density at $a=1$} & \parbox{2.5cm}{dimensionless\\Hubble's constant}\\ \hline
\end{tabular*}
\end{table}

\begin{figure}[h!]
\centering
\includegraphics[width=0.45\textwidth, angle=270]{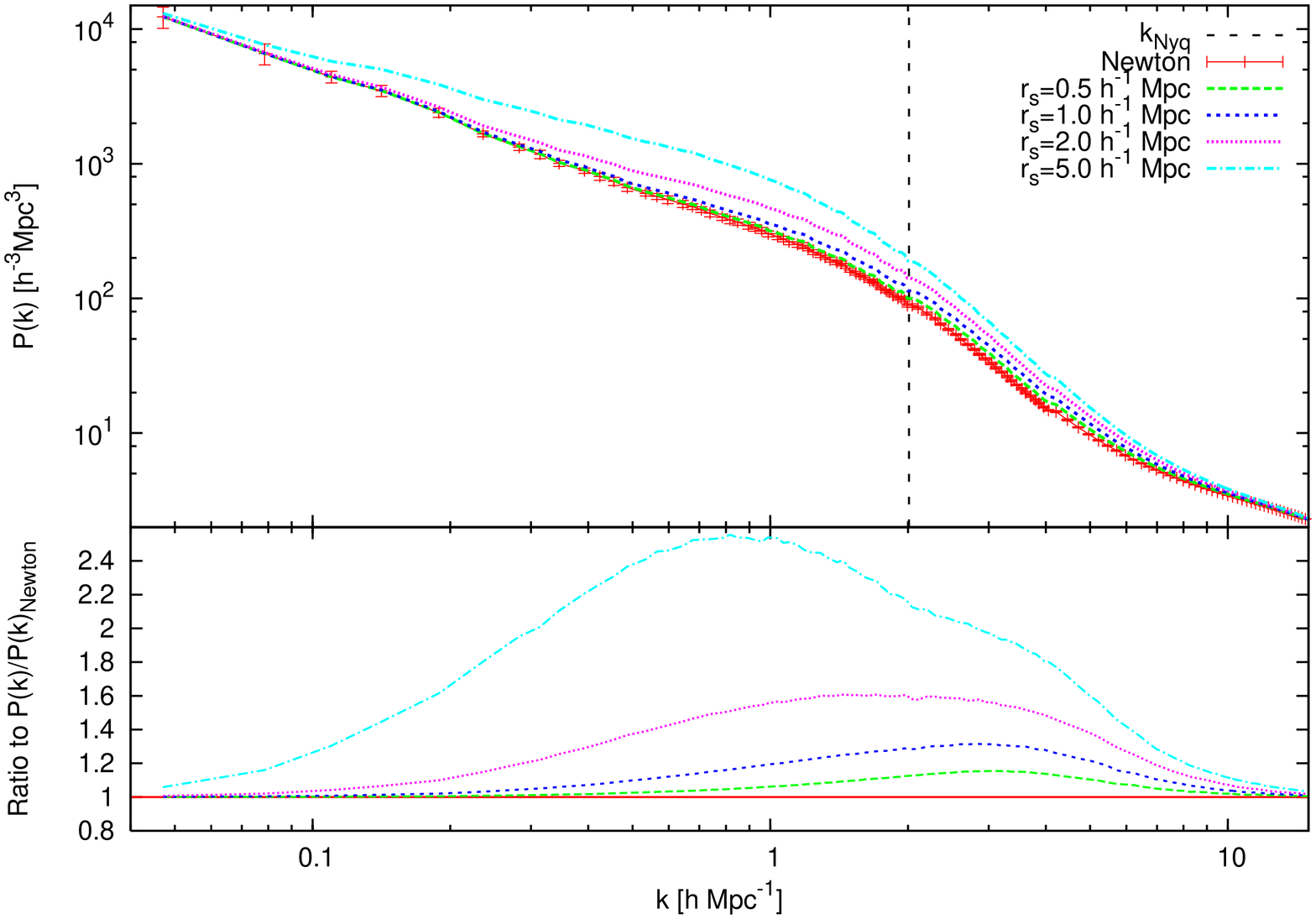}
\includegraphics[width=0.45\textwidth, angle=270]{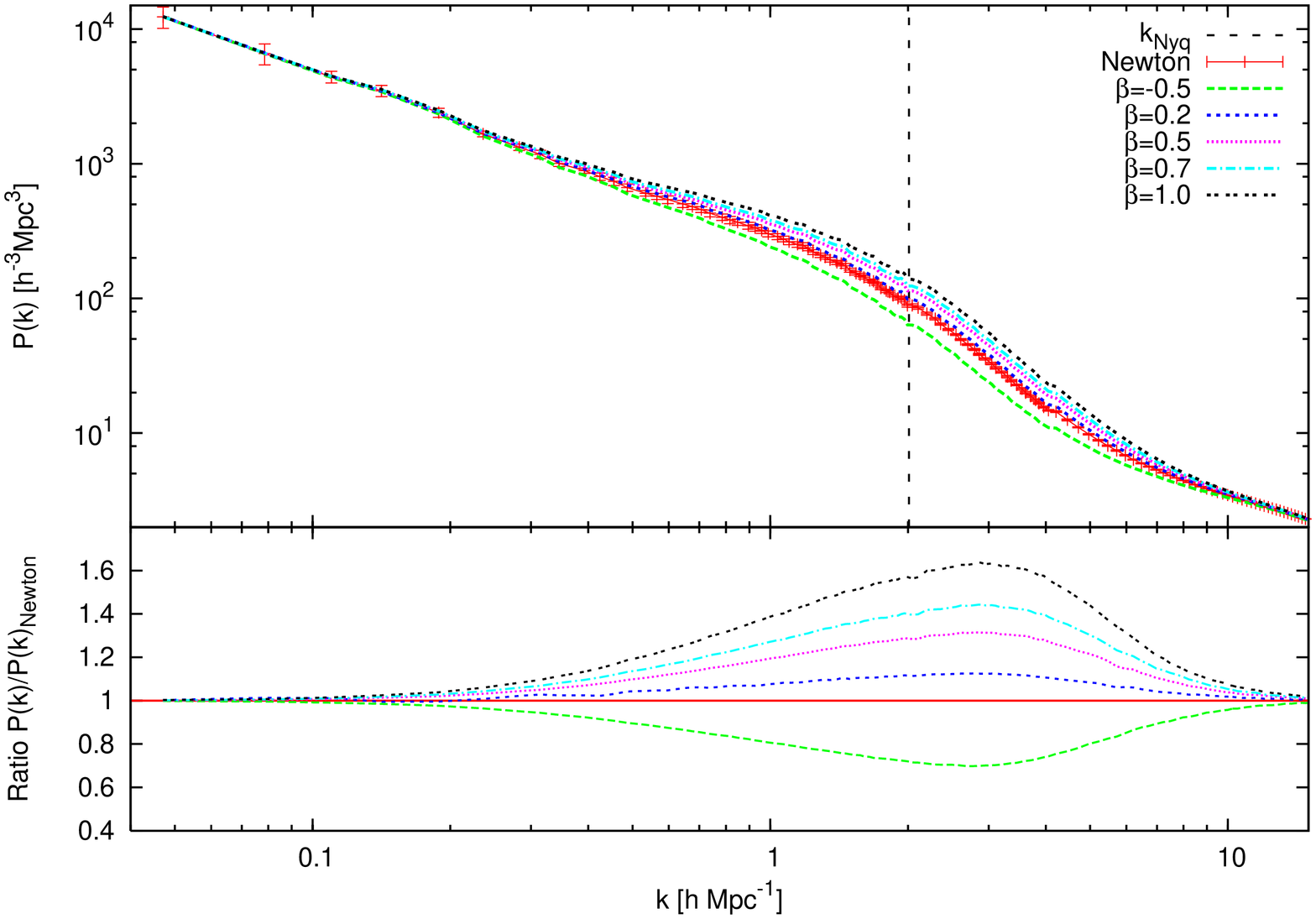}
\caption{Power spectrum of DM density perturbations. Top picture shows Pk(k) for a 
different values of $\beta$ parameter and for $r_s = 1.0 h^{-1}$Mpc fixed. Bottom picture illustrates power 
spectrum for fixed value of $\beta=0.5$ and for a different values of $r_s$. Lower panels of each picture shows relative 
ratio of measured power spectra to standard Newtonian gravity case. Black dotted line shows the Nyquist frequency. Data from simulation of $128^3$ DM particles in
a box width of $200h^{1}$ Mpc.}
\label{fig:1}
\end{figure}

\begin{figure}[h!]
\centering
\includegraphics[width=0.7\textwidth]{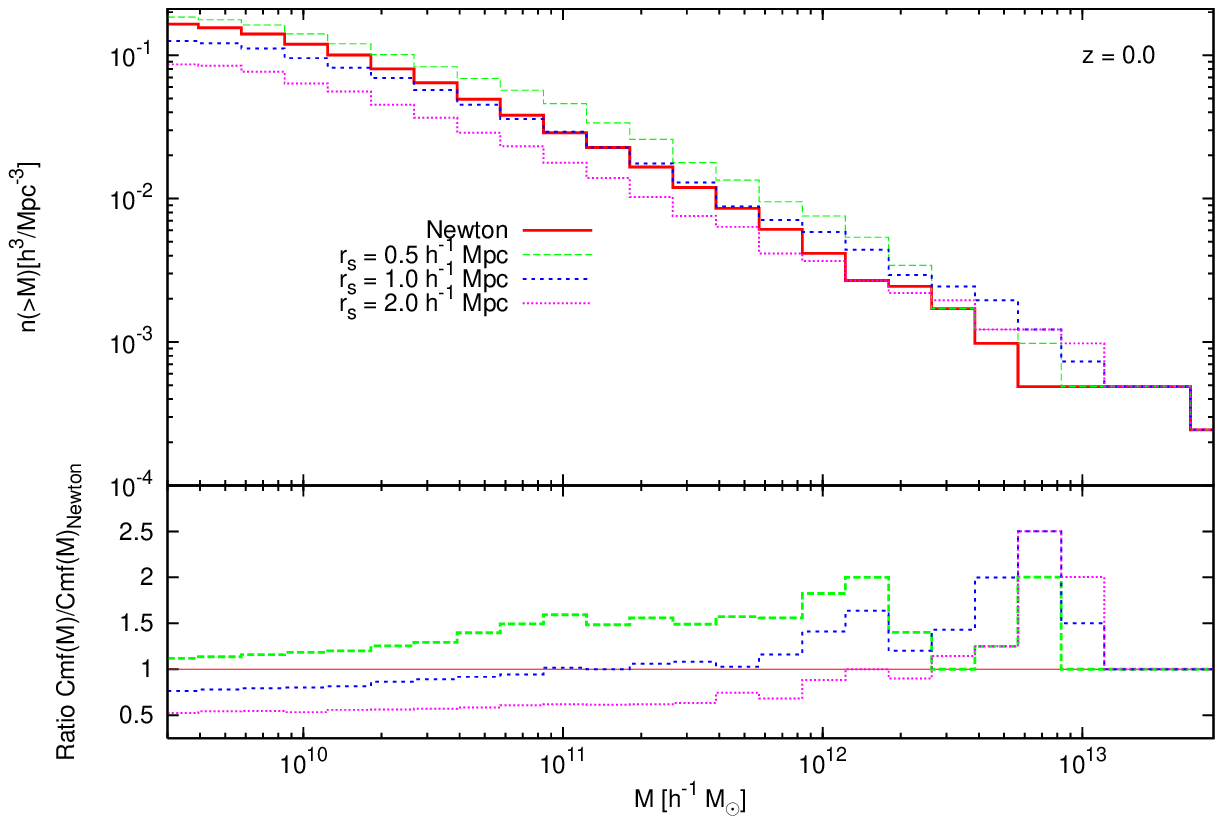}
\includegraphics[width=0.7\textwidth]{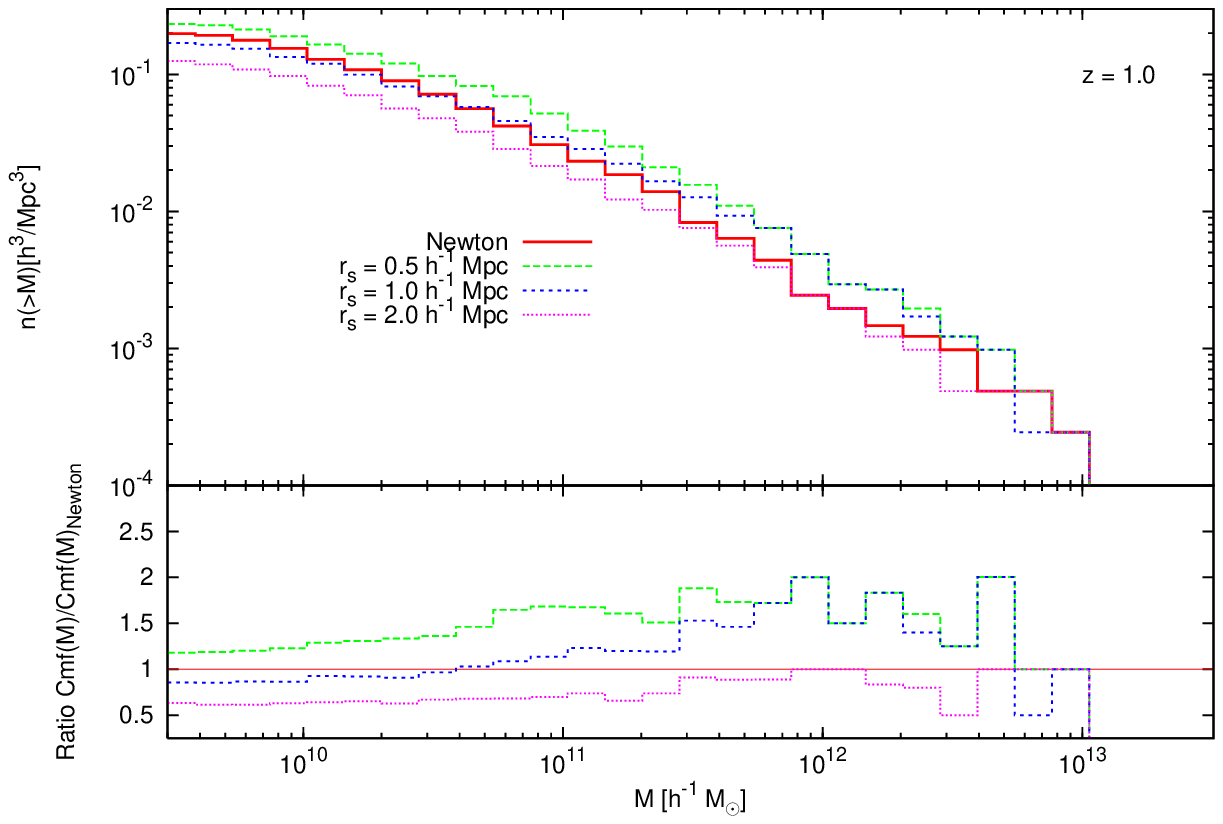}
\caption{CMF for redshifts z=0.0 (top figure) \& z=1.0 (bottom figure) ($\beta$ parameter is 1.0). Diagrams illustrates results 
from PM simulation of $123^3$ DM particles over $16 h^{-1}$ Mpc cube length}\label{fig:2}
\end{figure}

First we analyze the effect of scalar interaction on final power spectrum of DM density perturbations. 
Fig.\ref{fig:1} shows the calculated power spectra at redshift\footnote{redshift - describes a shift of a light 
wave towards longer wave tail caused by global expansion of the space. It is related to the evolution age of 
the Universe and the scale factor by $a={1\over 1+z}$} $z=0$ from simulations outputs. Each curve was obtained 
by averaging ten simulation runs for a given pair of parameters. For clarity we show error bars only for the Newtonian 
case. On the top picture we keep scalar power parameter $\beta$ fixed at 0.5 value. This allows us to analyze 
impact of changing value of screening length ($r_s$). It is clear to us that additional scalar interaction in 
DM substantial changes $P(k)$ only on intermediate range of scales related to $r_s$ ($k\sim 1/r_s$). Increasing 
the $r_s$ parameter increase the amplitude of deviation from Newtonian case as illustrated by the lower panel of the top figure. 
For $r_s= 5h^{-1}$Mpc case nearly all scales in power spectrum are affected.
Bottom picture shows $P(k)$ when we keep $r_s = 1h^{-1}$Mpc and change the $\beta$ parameter. 
We see that changing value of $\beta$ from -0.5 to 1.0 gently increase amplitude of fluctuations on scales 
corresponding to $k\sim 1h$Mpc$^{-1}$.

\begin{figure}[h!]
\centering
\includegraphics[width=0.7\textwidth]{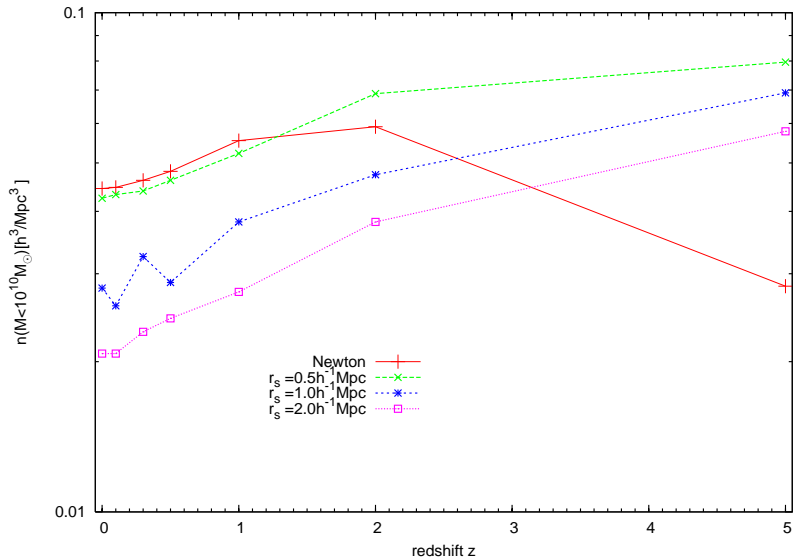}
\caption{Redshift evolution of the abundance of haloes with mass $M<10^{10}h^{-1}M_\odot$}\label{fig:3}
\end{figure}

Our PM simulations in the box of side length $200h^{-2}$Mpc are 
limited by force resolution of order $\sim 1$Mpc and mass resolution $\sim 10^{12} M_{\odot}$. 
To study the impact of scalar interaction of formation of structure on small-scale we have performed a series 
of simulation of $128^3$ particles in the box of side length $16h^{-1}$Mpc, this corresponds to the 
force resolution of order $\sim 300$kpc and mass resolution $\sim 10^{9} M_{\odot}$. In our analysis we included only
haloes with more then 20 particles which corresponds to minimal halo mass $M\sim 3.3\times10^9h^{-1}M_\odot$. In Fig.\ref{fig:2} 
we show calculated Cumulative halo Mass Function (CMF) for standard and modified gravity. CMF reflects the abundance 
of halos and it is normalized to number of halos of given mass $N(>M)$ per cubic megaparsec $h^{-3}$Mpc$^3$. We 
keep $\beta$ fixed to $1.0$ value and test the $r_s$ value. Top picture is calculated for current time $z=0$ the 
bottom picture is calculated at $z=1$. The lower panels presents relative ratio of CMF to the Newtonian case. These figures show that scalar 
interaction shifts the merging activity to the higher redshifts and towards higher masses. It is also clear that 
in the presence of scalar interaction with $r_s\geq1h^{-1}$Mpc the abundance of low-mass haloes is substantially 
lower than in the Newtonian case. We argue that this also indirectly implies that inter-galactic voids will be 
more empty in scalar model due to enhanced formation of small-scale structures. To emphasize this, Fig.\ref{fig:3} presents redshift
evolution of integral abundance of haloes with mass $M<10^{10}h^{-1}M_\odot$. Clearly scalar interaction models 
show much less evolution for $z\geq2$ than Newtonian model in this mass range. This indicates that small-scale structure growths 
earlier with scalar-interaction. Undoubtedly in our modified gravity models low-mass haloes are absorbed by high-mass cluster and galactic halos more efficiently than in Newtonian case. The fact that absorption of low-mass haloes starts on higher redshifts in scalar regime has obvious influence on finial
lower abundance of this haloes in scalar models.

We turn now to properties of the most massive halo (MMH) in each simulated model. Table \ref{table2} show general properties of the MMH for tested parameters. We show mass in virialized region, three-dimensional dispersion of velocity, velocity at virial radius, the virial radius, triaxiality parameter, ellipticities and the spin parameter. We calculate mass, radius and velocity at halo border where overdensity reaches 335 value. We define triaxiality parameter and ellipticies as:
\begin{eqnarray}
 T={a^2-b^2\over a^2-c^2},\\
e_1 = 1 - {c \over a}, \qquad e_2 = 1 - {b\over a},
\end{eqnarray}
where $a>b>c$ are eigenvalues of inertia tensor. For the spin parameter $\lambda$ we use the definition of Bullock et al.\cite{Bullock}:
\begin{equation}
 \lambda = {J\over \sqrt{2}M_{vir}v_{vir}r_{vir}} .
\end{equation}
The most dramatic effect of scalar forces that emerges from parameters in Table \ref{table2} is substantial excess of 3-dimensional velocity dispersion. The velocity dispersion in scalar models rise more rapidly than one could expect from increasing of the halo mass. We account this phenomenon as a purely modified gravitation effect. DM particles fall into the steeper potential wells than in Newtonian case and are gathered from larger co-moving distances during halo relaxation, thus they obtain more kinetic energy.

\begin{table}
\caption{General parameters of the most massive gravitationally bound object.}\label{table2}
\centering
\begin{tabular*}{\textwidth}{@{\extracolsep{\fill}}|c|c|c|c|c|c|c|c|c|c|}
\hline
model & $M_{vir}/(h^{-1}M_\odot)$ & $\sigma_v$ & $v_{cir}$ [km$\cdot$s$^{-1}]$ & $r_{vir}$ [kpc] & T & $e_1$ & $e_2$  &$\lambda$ \\ \hline
$\Lambda$ CDM & 4.64$\cdot 10^{13}$ & 570 & 522 & 732 & 0.86 & 0.55 & 0.44 & 0.013\\ \hline
$\beta = 1.0, r_s=0.5h^{-1}$Mpc & 4.62$\cdot 10^{13}$ & 784 &  521 & 731 & 0.88 & 0.55 & 0.46 & 0.03\\ \hline
$\beta = 1.0, r_s=1.0h^{-1}$Mpc & 5.41$\cdot 10^{13}$ & 867 &  550 & 770 & 0.83 & 0.49 & 0.38 & 0.018\\ \hline
$\beta = 1.0, r_s=2.0h^{-1}$Mpc & 6.8$\cdot 10^{13}$ & 973 &  593 & 831 & 0.78 & 0.52 & 0.37 & 0.032\\
\hline
\end{tabular*}
\end{table}

\begin{figure}[h!]
\centering
\includegraphics[width=0.45\textwidth,angle=270]{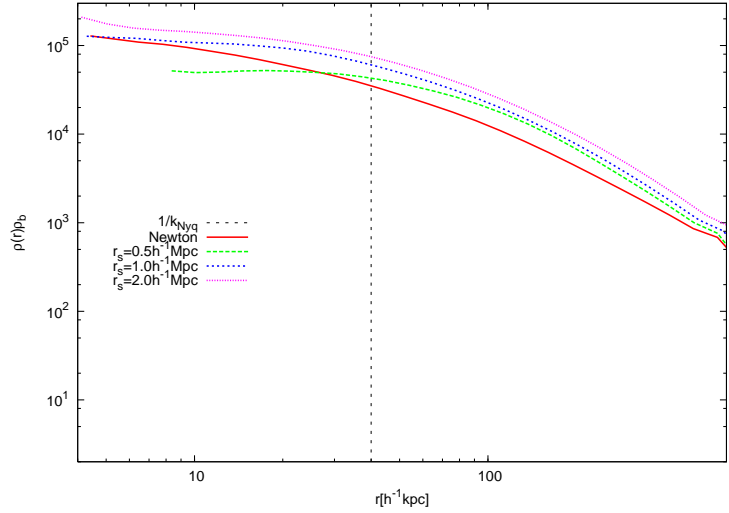}
\caption{Density profile for the most massive halo in scalar and Newtonian simulations. Vertical line indicate distance related to the Nyquist frequency of our simulations.}\label{fig:4}
\end{figure}

\begin{figure}[h!]
\centering
\includegraphics[width=0.45\textwidth,angle=270]{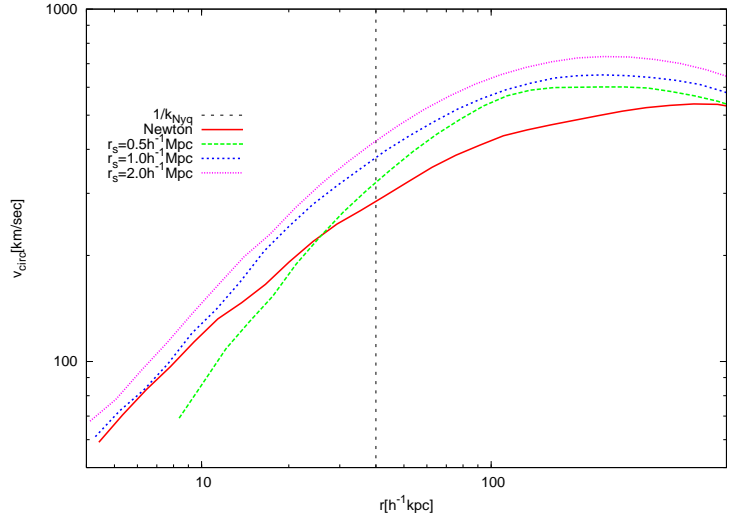}
\caption{Rotation curves of the MMH.}\label{fig:5}
\end{figure}

As our last but not least result in this paper we want to present calculated  MMH's density profiles (Fig.\ref{fig:4}) and velocity curves (Fig.\ref{fig:5}).
We want to outline the striking feature seen in density profiles, where a trace of solid density core can be found in halos simulated with the scalar interaction. Although this tantalizing feature would be most welcomed in context of astronomical observation we have to approach it with a caution. In our PM runs we do not resolve reliably scales smaller than $\approx 20\%$ of virial radius of the haloes. We have marked distance corresponding to the Nyquist frequency and, as can be seen, solid core feature appear below trusted distance threshold. More detailed and intense study of this effect is required to decide whether this property is real or just a ghost from numerics.

\section{Conclusions}
Our results show that  
the modification of standard gravity we consider can close the 
gap between some properties of standard $\Lambda$CDM cosmology simulations and today's high resolution astronomical 
observations. Deviations from the standard power spectrum are negligible, and they agree with large scale observations. 
On the other hand, our simulations of scalar interaction show desired clustering properties on small scales. 
While the court on the physical nature of dark matter particles is still out it 
is scientifically justified to probe various theoretical possibilities.
In this paper we have shown that the long-range 
scalar DM interaction has the potential to improve the standard cosmological model.

\section*{Acknowledgments}
Author acknowledges support from Polish Ministry of Science Grant No NN203 394234. Computer 
simulations used in this paper were conducted on the psk cluster at the Nicolaus Copernicus Astronomical Center. 
Author would like to thank Roman Juszkiewicz, Alexander Knebe, Pawel Ciecielag, Michal Chodorowski and Radek Wojtak for fruitful discussions.
Roman Juszkiewicz and Maciej Bilicki are acknowledged for reading carefully the manuscript.

\end{document}